\documentclass[supplement]{ptptex}

\title{%
\hfill{\normalsize%
\vbox{\hbox{\rm DPNU-04-08}  }}\\
\vspace{0.1cm}
Pion Velocity near the Chiral Phase Transition~\footnote{%
 Talk given at the YITP workshop on ``Nuclear Matter under 
 Extreme Conditions'' (Matter03), Dec. 1-3, 2003, Kyoto, Japan.
 This talk is based on the works done in 
 Refs.~1 and 2.
}
}

\author{
Chihiro \textsc{Sasaki}%
}

\inst{
Department of Physics, Nagoya University, 
Nagoya, 464-8602, Japan
}

\abst{
We focus on the pion velocity at the critical temperature
and study the quantum and hadronic thermal effects
based on the vector manifestation (VM).
We show the non-renormalization property on the pion velocity $v_\pi$,
which is protected by the VM, and estimate the value of $v_\pi$
to be close to the speed of light.
}

\begin{document}

\maketitle


\vspace*{-0.2cm}

An enhancement of dielectron mass spectra below $\rho / \omega$
resonance observed at CERN SPS~\cite{ceres} can be explained 
by the dropping mass of $\rho$ meson~\cite{LKB} following the 
Brown-Rho scaling~\cite{BR}.
There is a scenario which certainly requires the dropping mass
of the vector meson and supprots the Brown-Rho scaling,
named the vector manifestation (VM)~\cite{HY:VM} where the massless 
vector meson becomes the chiral partner of pion at the critical point.
By using the effective field theory based on the hidden local 
symmetry (HLS), which includes both pions and vector mesons as 
the dynamical degrees of freedom,
and performing the Wilsonian matching which is one of the methods to 
determine the bare theory from the underlying QCD,
the VM has been formulated~\cite{HY:VM}.
As pointed out in Ref.~\citen{VM:matter},
in the chiral phase transition in hot and/or dense matter,
the {\it intrinsic temperature and/or
density dependences of the parameters} of the HLS Lagrangian,
which are obtained by integrating out the high energy modes
(i.e., the quarks and gluons above the matching scale) in medium, 
play the essential roles to realize the chiral symmetry restoration
consistently.
Further, the intrinsic temperature and/or density dependences cause 
the effect of Lorentz symmetry breaking to the bare parameters.

In Ref.~\citen{sasaki}, 
starting from the bare HLS theory without Lorentz invariance,
it was proven that the non-renormalization property
on the pion velocity in hot matter, which is protected by the VM.
This can be understood based on the idea of chiral partners:
Away from $T_c$, 
the contribution from the longitudinal mode of thermal vector meson
$\rho_L$ is 
suppressed owing to the Boltzmann factor $\exp [-M_\rho / T]$, 
and only the thermal pion contributes to the pion velocity.
Then there exists the hadronic thermal correction to the pion velocity.
On the other hand,
when we approach the critical temperature,
the vector meson mass goes to zero due to the VM.
Thus $\exp [-M_\rho / T]$ is no longer the suppression factor.
As a result, the hadronic thermal correction to the pion velocity is absent
due to the exact cancellation between the contribution from pion and 
that from its chiral partner $\rho_L$.
Similarly, the quantum correction generated from the pion loop 
is exactly cancelled by that from the $\rho_L$ loop.
Accordingly we conclude
\begin{equation}
 v_\pi(T) = V_{\pi,{\rm bare}}(T)
 \qquad \mbox{for}\quad T \to T_c,
\label{phys=bare}
\end{equation}
i.e., {\it the pion velocity in the limit $T \to T_c$
receives neither hadronic nor quantum corrections due
to the protection by the VM.}
When one point in the subspace of the parameter $V_{\pi,{\rm bare}}$ 
is selected through the matching
procedure as done in Ref.~\citen{HKRS:pv},
namely the value of $V_{\pi,{\rm bare}}$ is fixed,
the present result implies that the point does not move in a subspace
of the parameters. 
Approaching the restoration point of chiral
symmetry, the physical pion velocity itself flows into the fixed
point.

As is
discussed in Ref.~\citen{HKRS:pv}, we should in principle evaluate
the matrix elements in terms of QCD variables only in order for
performing the Wilsonian matching, which is as yet unavailable from
model-independent QCD calculations.  Therefore, we make an
estimation by extending the dilute gas approximation adopted in
the QCD sum rule analysis in low temperature region to the
critical temperature with including all the light degrees of
freedom expected in the VM. In the HLS/VM theory, both the
longitudinal and transverse vector mesons become massless at the
critical temperature.
At the critical point, the longitudinal vector meson
couples to the vector current
whereas the transverse vector mesons decouple from the theory. 
Thus we assume that thermal
fluctuations of the system are dominated near $T_c$ not only by
the pions but also by the longitudinal vector mesons. 
We evaluate the thermal matrix elements of the non-scalar operators
in the OPE, by extending the thermal pion gas approximation employed
in Ref.~\citen{hkl} to the longitudinal vector mesons that figure
in our approach. 
Then the value of $V_{\pi,{\rm bare}}$ is estimated as
$V_{\pi,{\rm bare}}(T_c) = 0.83 - 0.99\,$.
It should be stressed that
the $bare$ pion velocity is expressed 
in terms of the OPE variables through the matching.

Finally thanks to the non-renormalization property 
given in Eq.~(\ref{phys=bare}),
we found that the pion velocity near $T_c$ is close to
the speed of light, $v_\pi (T) = 0.83 - 0.99\,.$
This is in contrast to the result obtaied from the chiral 
theory~\cite{SS},
where the relevent degree of freedom near $T_c$ is only the pion.
Their result is that the pion velocity becomes zero for $T \to T_c$.
Therefore from the experimental data,
we may be able to distinguish which picture is correct, 
$v_\pi \sim 1$ or $v_\pi \to 0$.

\section*{Acknowledgements}

I would like to thank
Professor Masayasu Harada, Doctor Youngman Kim and
Professor Mannque Rho
for many useful discussions and comments.
This work is supported in part by the 21st Century COE
Program of Nagoya University provided by Japan Society for the
Promotion of Science (15COEG01).

%

\end{document}